\documentclass{anstrans}
\title{Legendre Expansion for Scattering Anisotropy in Analytical 1D Multigroup $S_N$ Equations}
\author{Jilang Miao,$^{*1}$ and Miaomiao Jin$^{*}$}

\institute{
$^{*}$Department of Nuclear Engineering, The Pennsylvania State University, University Park, 16802 PA, USA
}
\footnotetext[1]{\footnotesize Corresponding author: Jilang Miao (jlmiao@psu.edu)}

\usepackage{graphicx} 
\usepackage{subfigure}

\usepackage{booktabs} 
\usepackage{multirow}
\usepackage{microtype} 
\usepackage{upgreek}
\usepackage{amsmath}
\usepackage{bbm}
\usepackage{algorithm}
\usepackage{algpseudocode}

\newcommand{\SN}{S$_N$~}

\expandafter\def\expandafter\normalsize\expandafter{%
    \normalsize%
    \setlength\abovedisplayskip{1pt}%
    \setlength\belowdisplayskip{1pt}%
    \setlength\abovedisplayshortskip{1pt}%
    \setlength\belowdisplayshortskip{1pt}%
}

\begin{document}
\section{Introduction}
The discrete ordinates method, commonly referred to as the \SN method~\cite{Car1953}, involves discretizing the particle transport equation in its differential form. The computation of particle fluxes relies on the direct evaluation of the transport equation at a finite set of discrete angular directions, referred to as ordinates.  
Additionally, quadrature relationships are employed to substitute integrals over angles, by converting them into summations over these discrete ordinates~\cite{hebert2009applied}.

Numerous efforts in the literature have concentrated on developing one-dimensional (1D) analytical transport solutions to meet diverse needs. However, the analytic solutions are typically confined to applications characterized by spatial homogeneity, angular isotropy (or linear scattering), and one energy group~\cite{siewert1965exact,de1990numerical,segatto1999one,ganapol2015response,WARSA2002851}. Recent advancements involve the vectorization of the \SN transport equation and seeking analytical solutions through matrix inversion ~\cite{wang2017new,wu2019SASN}. Although these methods can handle heterogeneous 1D problems, their applicability has been limited to one-group scenarios. Furthermore, they operate as fixed source solvers, necessitating the representation of the source term using piece-wise constant functions on a fine mesh for solving eigenvalue problems~\cite{wang2022}. Additionally, these approaches are constrained to matrices with real eigenvalues, precluding method acceleration such as redistributing more fission from the source term via Wielandt's shift~\cite{brown2007wielandt}.

To advance the capability of these methods~\cite{wang2017new,wu2019SASN,wang2022}, we previously developed an analytical solution for heterogeneous slab problems~\cite{analytical2A2G}. This solution, with closed-form expressions for $S_2$ and two energy groups, eliminates the need for power iteration to address eigenvalue problems. The matrix block-diagonalization procedures employed in this analytical approach facilitate the efficient treatment of complex eigenvalues. Subsequently, the method was expanded to tackle multigroup \SN equations in slab geometry~\cite{AMGSN2023}. This extension characterizes the solution within each grid through an expansion based on the eigensystem determined by neutron cross sections in the material. The expansion coefficients are determined by solving a linear system that incorporates continuity conditions at the interfaces and boundary conditions of the angular fluxes. The eigenvalues are obtained by seeking the root of the determinant of the boundary condition matrix.

Furthermore, we devised a fixed source solver for the multigroup \SN equations and applied it within the power iteration framework to handle eigenvalue problems. In the study presented in~\cite{AMGSNfixedsource}, power iteration was executed, assuming a piece-wise constant source on a fine mesh, while the fluxes were represented on a coarse mesh characterized by distinct materials. In a complementary work~\cite{AMGSNcoarsemesh}, a coarse mesh iteration method was developed, wherein both flux and source terms are expanded based on the eigensystem determined by material cross sections. This method achieves accelerated computation while maintaining the same level of accuracy.

However, a common assumption in above-mentioned methods ~\cite{AMGSN2023,AMGSNfixedsource,AMGSNcoarsemesh} is that the scattering matrix is in the form of $\Sigma_{s, g^{\prime} n^{\prime} \rightarrow g n}$, where $g$ and $g^{\prime}$ are indices for energy groups, and $n$ and $n^{\prime}$ are indices for discrete angles. In the case of isotropic scattering, obtaining the matrix is straightforward from the scattering cross section without dependence on $n$ and $n^{\prime}$. However, for anisotropic scattering, generating such cross sections is not feasible. Anisotropic scattering is typically represented using Legendre expansion~\cite{stacey2018nuclear,wang2021rattlesnake}. In this study, we incorporate scattering anisotropy into the 1D analytical multigroup \SN equations using Legendre expansions. We demonstrate its accuracy through a comparison with Monte Carlo (MC) reference on a heterogeneous slab problem derived from a typical pincell.

\section{Methodologies}
For a given number of energy groups, denoted as $g=1,...,G$, 
and a quadrature set $\left . \{\mu_n,\omega_n\} \right | _{n=1,...,N}$, 
the transport equation for the angular flux $\psi_{g,n}$ is expressed in Eq~\ref{eq::sn1d}. 
\begingroup
\footnotesize
\begin{equation}
\begin{aligned}
& \mu_n \frac{\partial}{\partial x} \psi_{g,n}(x)+  \Sigma_{t, g} \psi_{g,n}(x)= 
   \sum_{n^{\prime}, g^{\prime}} \omega_{n \prime} \Sigma_{s, g^{\prime} n^{\prime} \rightarrow g n} \psi_{g^{\prime}, n^{\prime}}(x) \\
&+\frac{1}{k_{eff}}  \sum_{n^{\prime}, g^{\prime}} \omega_{n^{\prime}}  \nu \Sigma_{f, g^{\prime} n^{\prime} \rightarrow g n} \psi_{g^{\prime}, n^{\prime}}(x) 
\label{eq::sn1d}
\end{aligned}
\end{equation}
\endgroup
Since it is not practical to generate multigroup cross sections in the form of
$\Sigma_{s, g^{\prime} n^{\prime} \rightarrow g n}$,
we rewrite the scattering term as a function of Legendre moments.
The scattering rate $S$ from group $g'$ to $g$ with scattering cosine $\mu$ is conventionally written as Eq.~\ref{eq::sca l},
\begingroup
\footnotesize
\begin{equation}
\begin{aligned}
  & ~~~ S_{g^{\prime}\rightarrow g}(x,\mu) \\  &= 
  \sum_{l=0}^L\frac{2l+1}{2}\Sigma_{s,g^{\prime}\rightarrow g,l}P_l(\mu)\phi_l(x) \\
&= 
  \sum_{l=0}^L\frac{2l+1}{2}\Sigma_{s,g^{\prime}\rightarrow g,l}P_l(\mu) \int d\mu^{\prime} P_l(\mu^{\prime}) \psi_{g^{\prime}}(x,\mu^{\prime}) \\
&= 
  \sum_{l=0}^L\frac{2l+1}{2}\Sigma_{s,g^{\prime}\rightarrow g,l}P_l(\mu) \sum_{n^{\prime}} \omega_{n^{\prime}} P_l(\mu_{n^{\prime}}) \psi_{g^{\prime},n^{\prime}}(x) 
\end{aligned}\label{eq::sca l}
\end{equation}
\endgroup
Here, we replace the integral in the definition of $\phi_l(x)$ with
the sum over \SN quadrature sets. 
With $\mu=\mu_n$ in Eq.~\ref{eq::sca l} and plugging into Eq~\ref{eq::sn1d}, we can organize the cross-sections and quadrature sets into matrices ~\cite{AMGSN2023}, 
and hence Eq~\ref{eq::sn1d} can be written in matrix form as,
\begingroup
\footnotesize
\begin{equation}
    \partial_x \Psi(x) = A \Psi(x) 
    \label{eq::sn1d matrix}
\end{equation}
\endgroup
where
\begingroup
\footnotesize
\begin{equation}
  A = \frac{F}{k_{eff}} + S - T
  \label{eq::TFStoA}
\end{equation}
\endgroup
Herein, $F$, $S$, and $T$ are the respective fission, scattering, and total cross sections multiplied by \SN quadrature set parameters  $\left . \{\mu_n,\omega_n\} \right | _{n=1,...,N}$. 
Specifically, for scattering, we have 
\begingroup
\footnotesize
\begin{equation}
  S_{gN+ n, g^{\prime}N+ n^{\prime}  } = 
  \frac{\omega_{n^{\prime}} } {\mu_n } 
  \sum_{l=0}^L\frac{2l+1}{2} P_l(\mu_n) \Sigma_{s,g^{\prime}\rightarrow g,l} P_l(\mu_{n^{\prime}})   
  \label{eq::Smat l}
\end{equation}
\endgroup
With this formulation, we can treat scattering anisotropy to arbitrary Legendre order.
Then the multigroup \SN eigenvalue problems in heterogeneous slab systems can be solved via either the non-iterative determinant root solver~\cite{AMGSN2023} or the iterative methods~\cite{AMGSNfixedsource,AMGSNcoarsemesh}. 

\section{Results}
In this section, we test the method on a heterogeneous slab system where multigroup cross sections are generated from a pincell.
Results including $k_{eff}$ , scalar fluxes and angular fluxes from $S_4$, $S_8$, $S_{16}$, and $S_{32}$ will be compared with a MC reference.

\subsection{Description of the test case}
We consider a  pincell with UO$_2$ fuel, helium gap, zircaloy cladding and borated water.
The pincell has a pitch of 1.323 cm and length of 30 cm. Two
borated water regions with 2.5 cm thickness are on the ends of the fuel, respectively.
Boundary conditions are vacuum axially and reflective radially.
Two-group cross-sections are generated with OpenMC~\cite{romano2013openmc,boyd2019multigroup}
with scattering Legendre moments of order 4 ($P4$).
The pincell is homogenized to a slab problem with the core of length 30cm and two reflector regions of length 2.5cm on both ends. 
The cross sections for the core and reflector materials are shown in Table~\ref{tab::xs}.

\begingroup
\begin{table}[htb]
\scriptsize
  \centering
  \caption{Cross-section parameters.}
  \begin{tabular}{llr}\toprule
      &  Core      & Reflector
    \\ \midrule
  $\Sigma_{t,1}$ & 6.8294e-01 & 8.9176e-01 \\
  $\Sigma_{t,2}$ & 2.0658e+00 & 3.0361e+00 \\
  $\Sigma_{s,0,1\rightarrow 1}$ & 6.4870e-01 & 8.4530e-01 \\
  $\Sigma_{s,0,1\rightarrow 2}$ & 2.5869e-02 & 4.6078e-02 \\
  $\Sigma_{s,0,2\rightarrow 1}$ & 4.2114e-04 & 2.8498e-04 \\
  $\Sigma_{s,0,2\rightarrow 2}$ & 1.9696e+00 & 3.0181e+00 \\
  $\Sigma_{s,1,1\rightarrow 1}$ & 3.2525e-01 & 5.0694e-01 \\
  $\Sigma_{s,1,1\rightarrow 2}$ & 7.7637e-03 & 1.4061e-02 \\
  $\Sigma_{s,1,2\rightarrow 1}$ & 2.2069e-04 & 2.0111e-04 \\
  $\Sigma_{s,1,2\rightarrow 2}$ & 4.4646e-01 & 6.6720e-01 \\
  $\Sigma_{s,2,1\rightarrow 1}$ & 1.3329e-01 & 2.0454e-01 \\
  $\Sigma_{s,2,1\rightarrow 2}$ & -2.5799e-03 & -4.6366e-03 \\
  $\Sigma_{s,2,2\rightarrow 1}$ & 1.3804e-04 & 1.2919e-04 \\
  $\Sigma_{s,2,2\rightarrow 2}$ & 9.3323e-02 & 1.2844e-01 \\
  $\Sigma_{s,3,1\rightarrow 1}$ & 1.1392e-02 & 1.2657e-02 \\
  $\Sigma_{s,3,1\rightarrow 2}$ & -3.0492e-03 & -5.4962e-03 \\
  $\Sigma_{s,3,2\rightarrow 1}$ & 7.1796e-05 & 6.7691e-05 \\
  $\Sigma_{s,3,2\rightarrow 2}$ & 2.1589e-02 & 2.8195e-02 \\
  $\Sigma_{s,4,1\rightarrow 1}$ & -1.4437e-02 & -2.8207e-02 \\
  $\Sigma_{s,4,1\rightarrow 2}$ & -6.6600e-04 & -1.1974e-03 \\
  $\Sigma_{s,4,2\rightarrow 1}$ & 2.2672e-05 & 2.2173e-05 \\
  $\Sigma_{s,4,2\rightarrow 2}$ & -2.7343e-03 & -5.7950e-03 \\
  $\nu\Sigma_{f,1}$ & 6.0427e-03 & 0.0000e+00 \\
  $\nu\Sigma_{2,2}$ & 1.5343e-01 & 0.0000e+00 \\
  $\chi_1$ & 1.0000e+00 & 0.0000e+00 \\
  $\chi_2$ & 0.0000e+00 & 0.0000e+00 \\
\bottomrule    
\end{tabular}
  \label{tab::xs}
\end{table}
\endgroup

\subsection{Scattering anisotropy}
In addition to obtaining the multigroup cross sections, the continuous energy Monte Carlo simulation on the original pincell yields its eigenvalue at 1.13604 $\pm$ 2pcm. Subsequently, we perform a multigroup MC simulation to generate a reference for the slab problem, utilizing the cross-sections as listed in Table~\ref{tab::xs}. The eigenvalue of the slab is determined to be 1.16055 $\pm$ 1pcm. Assuming isotropic scattering and neglecting the higher Legendre moments, the eigenvalue of the slab is calculated to be 1.24953 $\pm$ 2pcm. The $k_{eff}$ values are summarized in Table~\ref{tab::keff mc}. Within the total $11349$ pcm error of the isotropic model, the discrepancy between isotropic and $P4$ scattering accounts for $8895$ pcm $k_{eff}$ error, while the remaining $2454$ pcm error is attributed to radial homogenization, the two-group approximation, and higher-order scattering moments.
\begingroup
\begin{table}[htb]
\scriptsize
  \centering
  \caption{$k_{eff}$ reference of different models.}
  \begin{tabular}{llllr}\toprule
\multicolumn{3}{c}{Model Configuration}     & \multirow{ 2}{*}{$k_{eff}$}  &  diff  \\ 
Energy  & Scatter &  Geometry   &                              &  (pcm) \\ 
    \midrule
\multicolumn{2}{l}{ENDF/B-VII.1}  & 3D & 1.13604 $\pm$ 2pcm &  \\
\midrule
2 group &isotropic & slab     & 1.24953 $\pm$ 2pcm & -11349  \\
2 group &$P4$      & slab     & 1.16055 $\pm$ 1pcm & -2454 \\
    \bottomrule
\end{tabular}
  \label{tab::keff mc}
\end{table}
\endgroup

\begin{figure}[htbp]
\centering
\subfigure[]
{\includegraphics[width=0.21\textwidth]{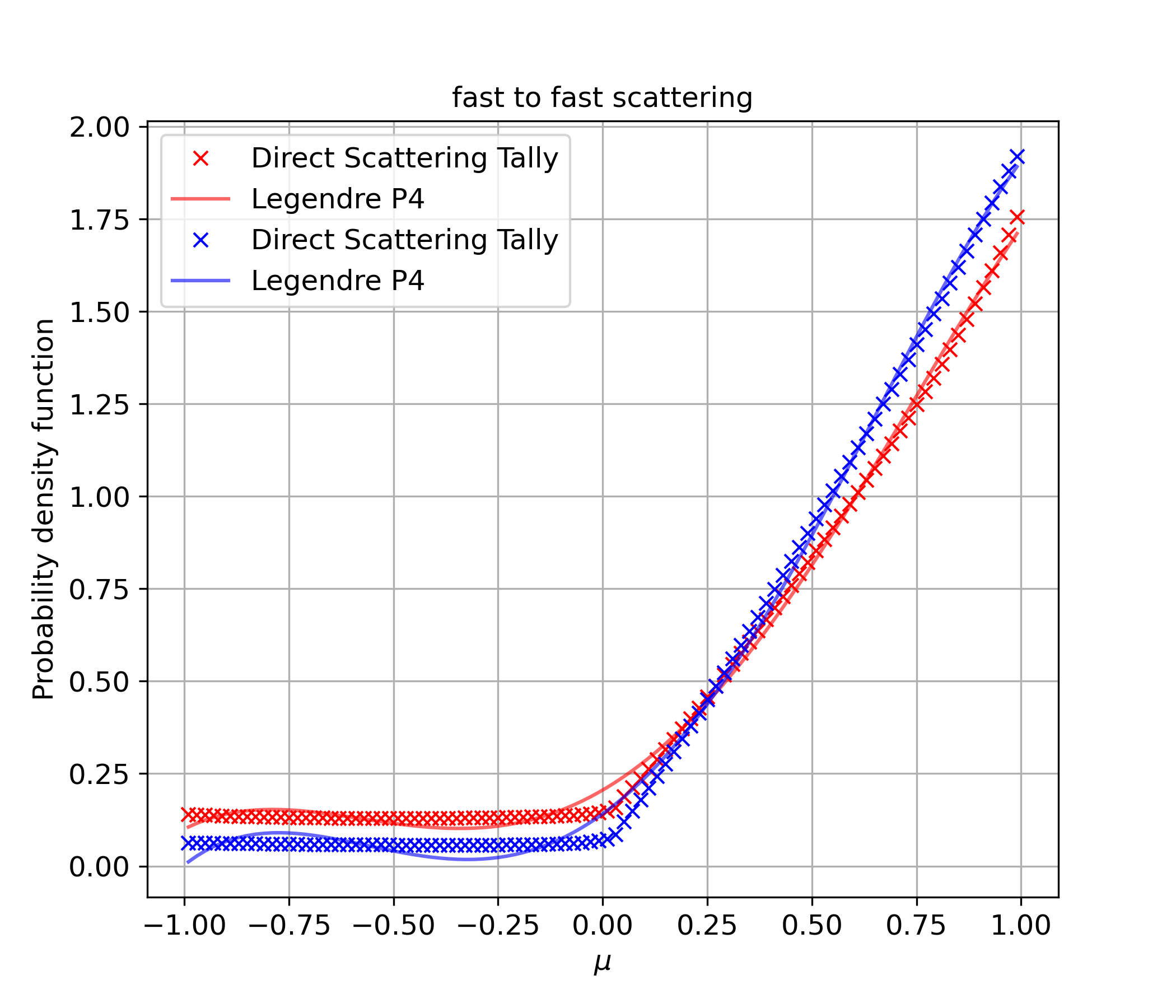}
} 
\subfigure[]
{\includegraphics[width=0.21\textwidth]{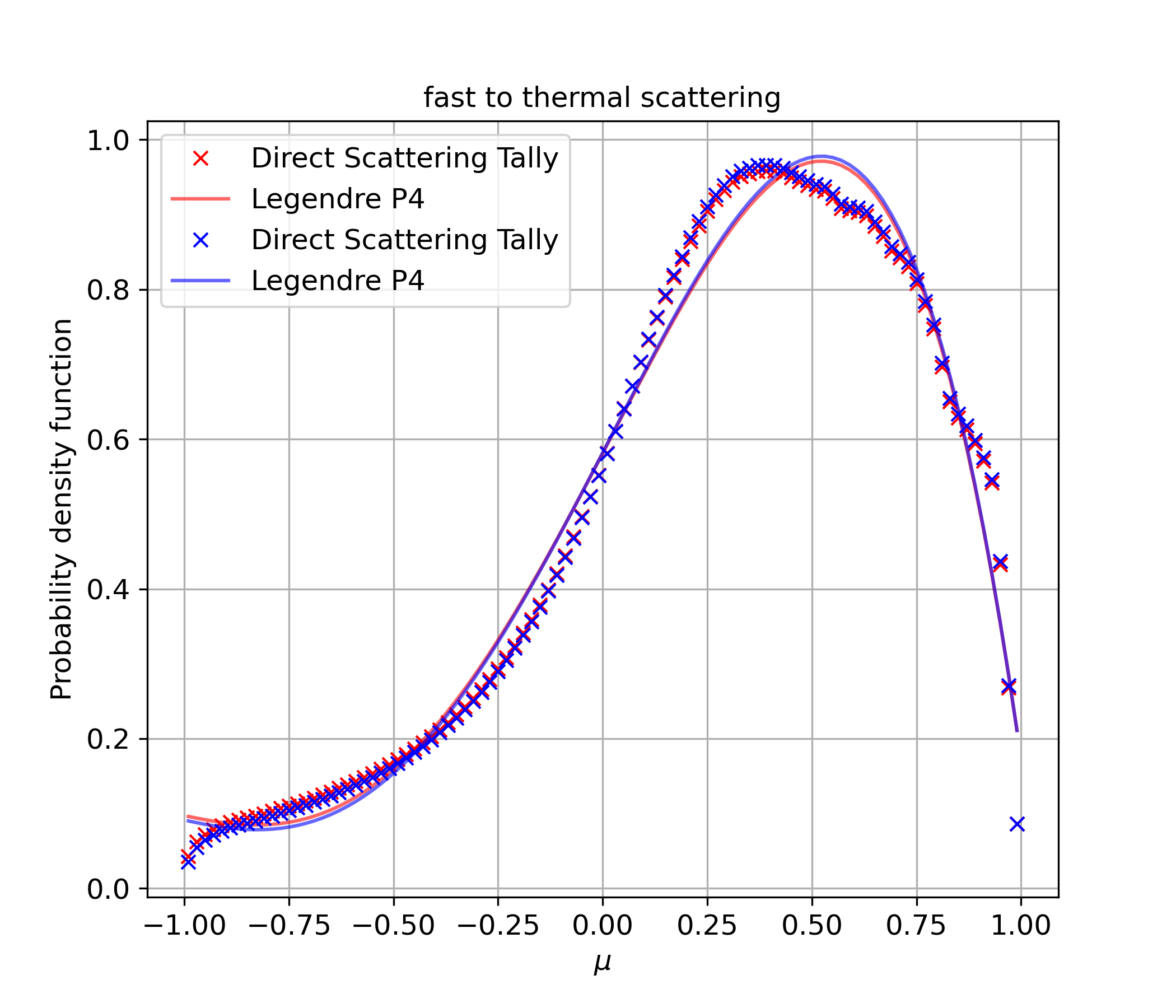}
} 
\subfigure[]
{\includegraphics[width=0.21\textwidth]{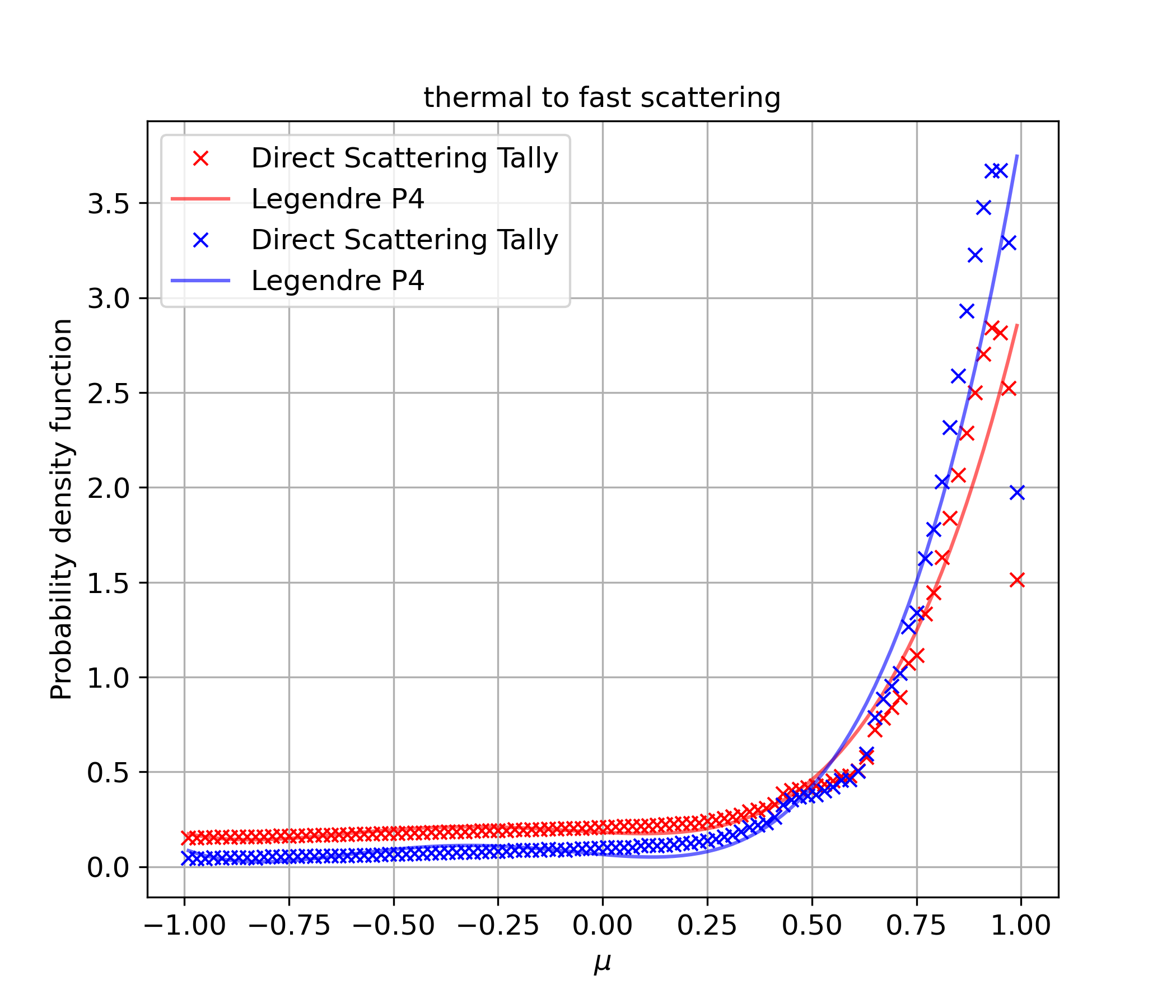}
} 
\subfigure[]
{\includegraphics[width=0.21\textwidth]{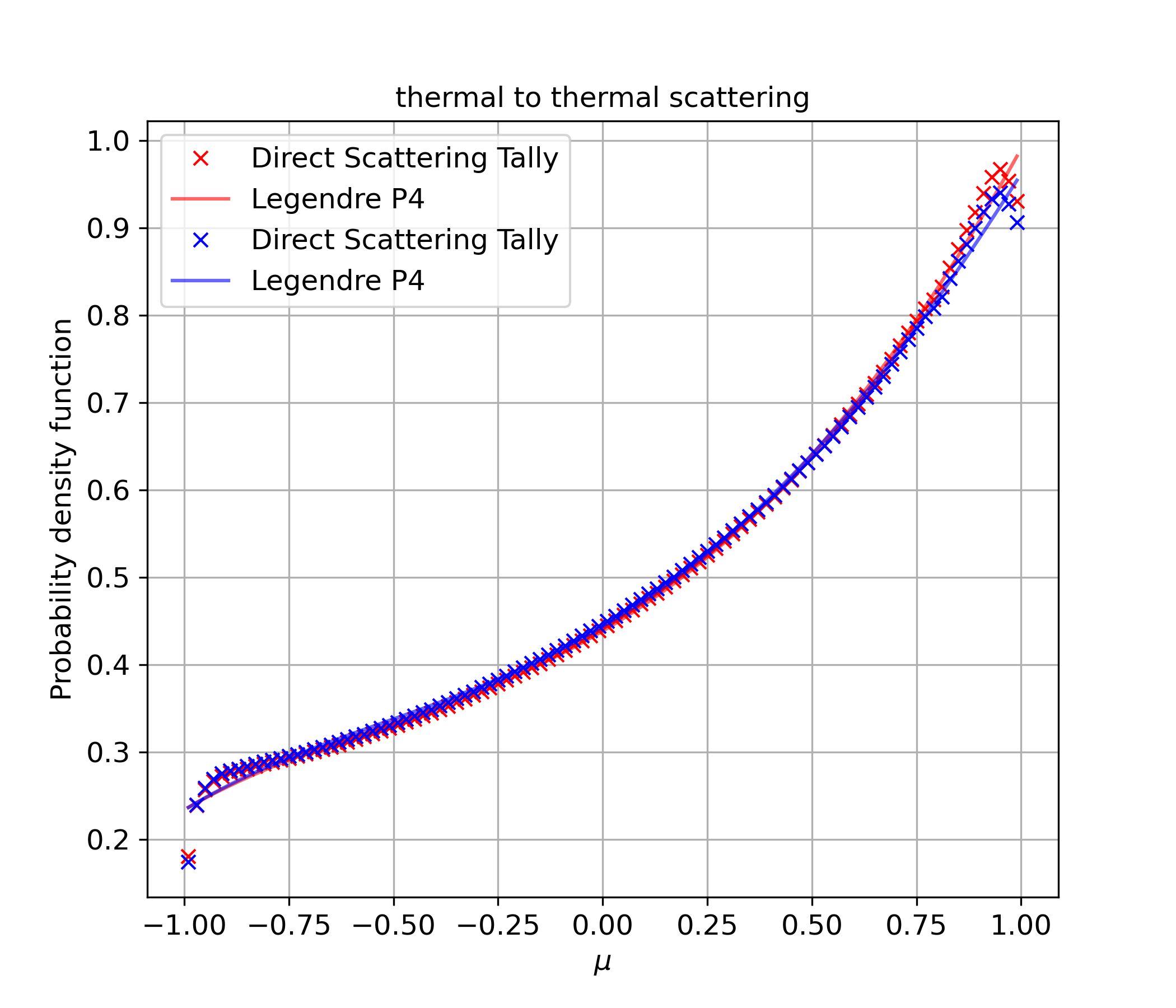}
} 
\caption{\scriptsize Scattering Angle distribution. The crosses are based on tallies from the continuous MC simulation,
and the solid curves are from $P4$ Legendre expansion.  }
\label{fig::sigs mu}
\end{figure}

Scattering anisotropy is also demonstrated by the scattering angle distribution for each
incoming and outgoing energy group in core and reflector.
The scattering angle distribution from both MC tallies and Legendre expansions are given in Fig.~\ref{fig::sigs mu}.
Here, the scattering distributions are calculated using Eq.~\ref{eq::plt pmu} with cross sections from Table~\ref{tab::xs}.
The consistency between MC tallies and Legendre expansion demonstrates that $P4$ is accurate enough to capture the anisotropy of the problem. In particular, the probability density function of the scattering angle clearly shows the scattering is not isotropic.

\begingroup
\footnotesize
\begin{equation}
\sum_{l=0}^4 P_l(\mu) \Sigma_{s,g^{\prime}\rightarrow g,l}
\label{eq::plt pmu}
\end{equation}
\endgroup

\subsection{Numerical Results}
The reference for the test case is derived using the multigroup mode of OpenMC. The simulation tracks $2\times 10^6$ neutrons per generation. Neutrons are simulated for $200$ inactive generations, and tallies are collected for the subsequent $1000$ active generations to compute scalar fluxes, angular fluxes, and $k_{eff}$. Fluxes are tallied on a spatially uniform mesh with a size of $700$ for each energy group. Additionally, angular fluxes are tallied over a specific polar angle range corresponding to the \SN quadrature set.

We then solve the eigenvalue of the heterogeneous slab with the analytical multigroup \SN methods introduced in ~\cite{AMGSN2023,AMGSNfixedsource,AMGSNcoarsemesh}.
All the methods return the eigenvalue and expansion coefficients for angular fluxes.
The expansion coefficients are then used to evaluate the angular fluxes on the same $700$ spatial mesh to compare spatially-dependent flux values. The results from these methods on  $S_4$, $S_8$, $S_{16}$, $S_{32}$ are compared in Table~\ref{tab::keff}.
The ``Determinant Root Solver'' method builds the linear system from boundary conditions and interface angular flux continuity conditions on the coarse mesh~\cite{AMGSN2023}.
It determines the eigenvalue as the root of the determinant of the boundary condition matrix and solves the angular flux coefficients as the null space of the boundary condition matrix.
The ``Coarse Mesh Iteration'' method represents both angular flux and source term on the coarse mesh and updates the coefficients via power iteration~\cite{AMGSNcoarsemesh}.
The ``Fine Mesh Iteration'' method represents the source term as piece-wise function on the fine mesh, solves the angular flux expansion coefficients on the coarse mesh and updates the coefficients via power iteration~\cite{AMGSNfixedsource}. 
The values under `mesh' column correspond to the grid number in each region with a format (reflector-core-reflector); for the fine mesh method, there is a fourth number which is the grid size for the source term.


Table~\ref{tab::keff} distinctly illustrates the convergence of the solution toward the Monte Carlo reference, reducing from $-48$ pcm for $S_4$ to less than $1$ pcm for $S_{32}$. Additionally, it is noteworthy that all the analytical methods evaluated exhibit eigenvalue differences below $0.1$ pcm.
\begingroup
\begin{table}[htb]
\scriptsize
  \centering
  \caption{$k_{eff}$ from \SN compared with MC reference.}
  \begin{tabular}{lllr}\toprule
\multicolumn{2}{l}{Method}      & $k_{eff}$      & error (pcm) \\ \midrule 
\multicolumn{2}{l}{MC reference} & 1.160548 $\pm$ 1.5pcm &  \\  
\midrule

   \multicolumn{4}{l}{Determinant Root Solver}  \\
      order & mesh     &          &  \\
  $S_4$     & (1-4-1)  & 1.160069 & -47.9 \\
  $S_8$     & (1-9-1) & 1.160455 & -9.3 \\
  $S_{16} $ & (2-18-2) & 1.160534 & -1.4 \\
  $S_{32}$  & (3-30-3) & 1.160552 & 0.4 \\ \midrule 
   \multicolumn{4}{l}{Coarse Mesh Iteration}  \\
      order & mesh     &          &  \\
  $S_4$     & (1-4-1) & 1.160069 & -47.9 \\
  $S_8$     & (1-9-1)& 1.160455 & -9.3 \\
  $S_{16} $ & (2-18-2) & 1.160534 & -1.4 \\
  $S_{32}$  & (2-36-2)& 1.160552 & 0.4 \\ \midrule 
   \multicolumn{4}{l}{Fine Mesh Iteration}  \\
      order & mesh     &          &  \\
  $S_4$    & (1-4-1+696)& 1.160069 & -47.9 \\
  $S_8$    & (1-9-1+693)& 1.160455 & -9.3 \\
  $S_{16}$ & (2-18-2+686)& 1.160534 & -1.4 \\
  $S_{32}$ & (2-36-2+680)& 1.160552 & 0.4 \\ 
    \bottomrule
\end{tabular}
  \label{tab::keff}
\end{table}
\endgroup

\begin{figure}[htbp]
\centering
\subfigure[$S_4$]
{\includegraphics[width=0.4\textwidth]{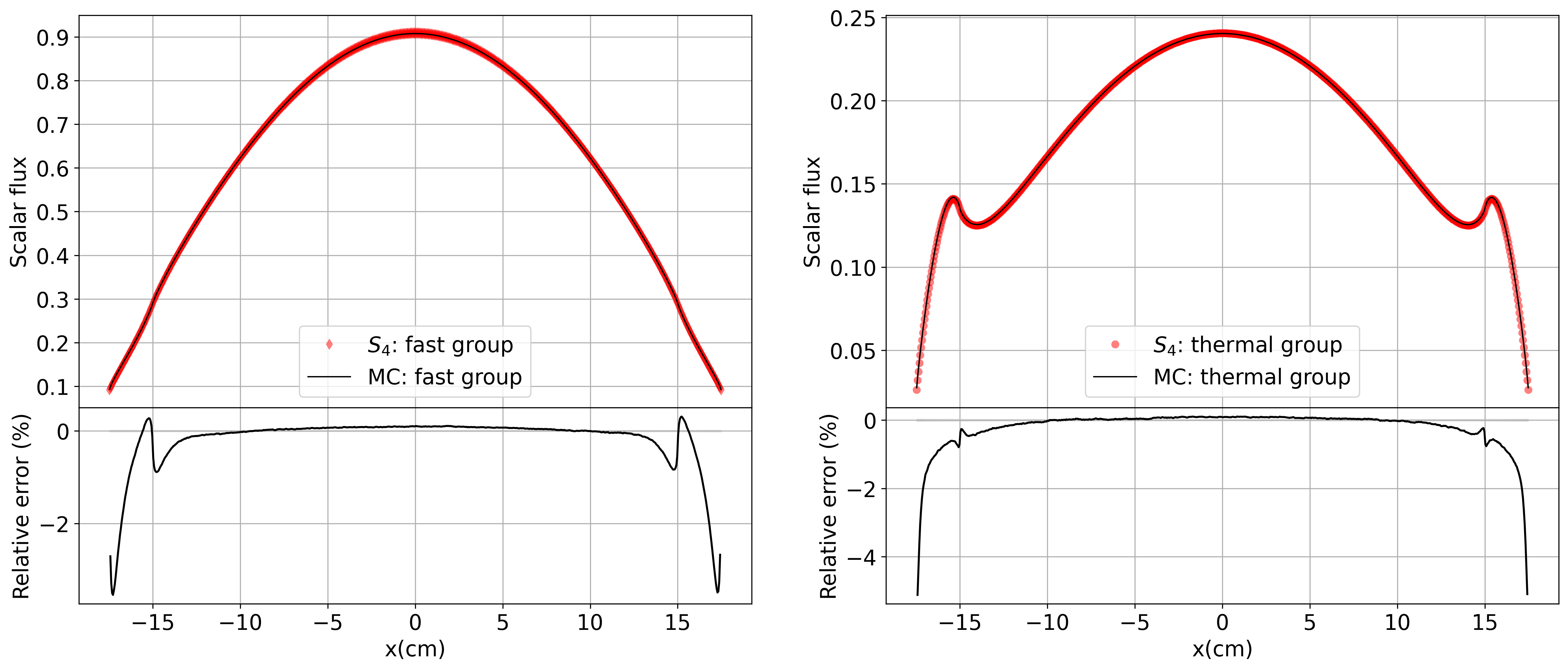}
  \label{fig::phi S4} 
} 
\subfigure[$S_8$]
{\includegraphics[width=0.4\textwidth]{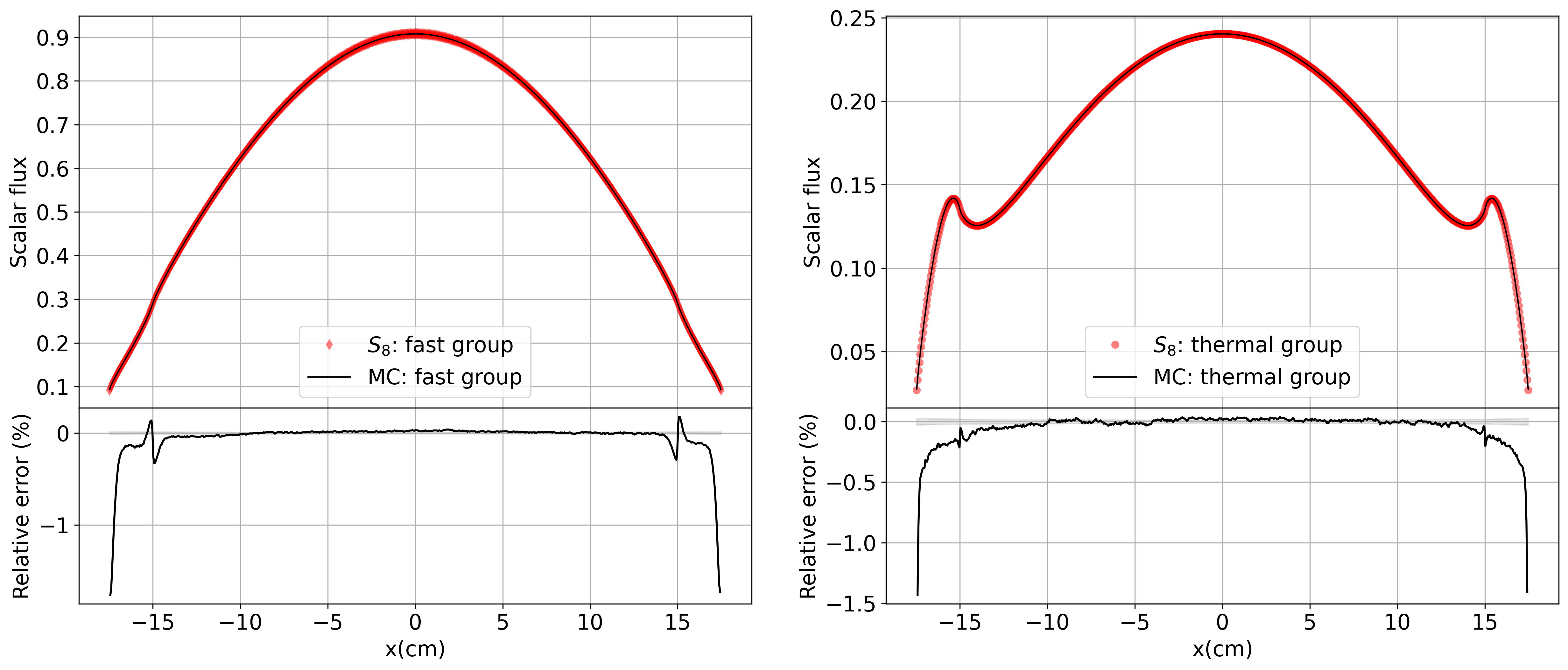}
  \label{fig::phi S8} 
} 
\subfigure[$S_{16}$]
{\includegraphics[width=0.4\textwidth]{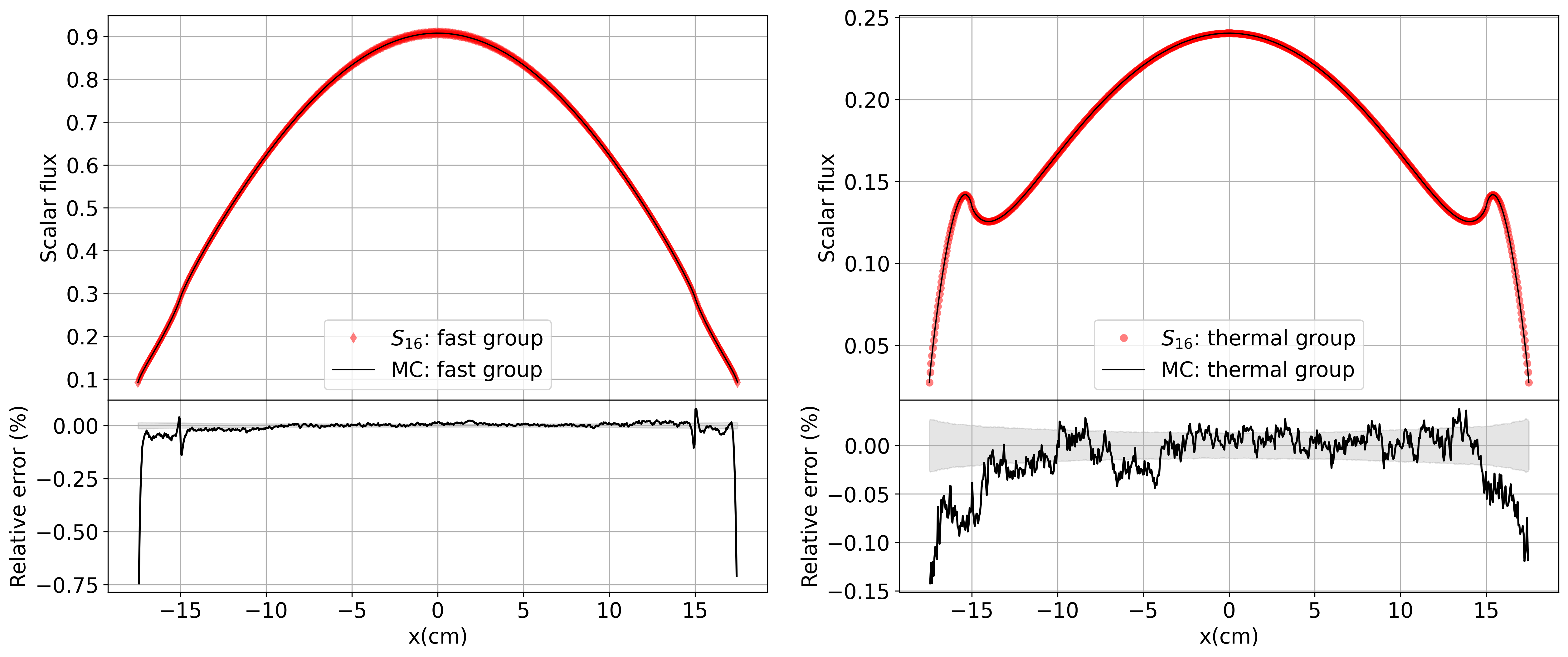}
  \label{fig::phi S16} 
} 
\subfigure[$S_{32}$]
{\includegraphics[width=0.4\textwidth]{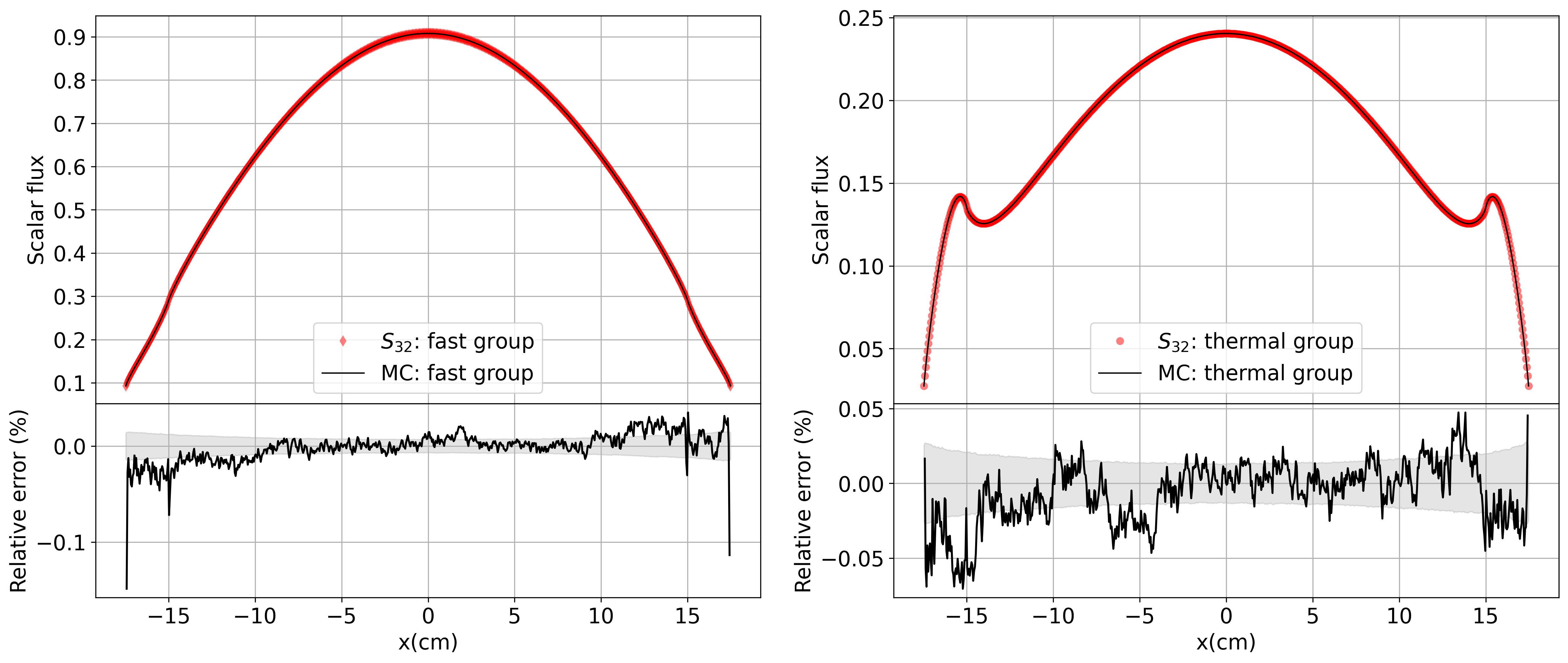}
  \label{fig::phi S32} 
} 

\caption{\scriptsize \SN scalar fluxes compared with MC.
  For each figure, the upper part plots MC reference in solid curve and \SN results with diamond symbol for fast group and circle symbol for thermal group, and the lower part plots the pointwise relative error (\%) between \SN and MC.
  The standard deviation of each tally $T$ from MC is shown with the shading area between $\pm 100\times \frac{\sigma_{\bar{T}}}{\bar{T}}$.}
\label{fig::phi comp}
\end{figure}

Next, we proceed to compare the scalar fluxes, as illustrated in Fig.~\ref{fig::phi comp}, 
which includes results from $S_4$, $S_8$,$S_{16}$ and $S_{32}$.
For each order, the scalar fluxes from \SN and MC are compared for fast and thermal groups, respectively.
Within each subfigure, the upper plots show the accurate match of the scalar fluxes,
while the bottom plots depict the point-wise relative error in percentage between \SN and MC references. 
Notably, as the orders increase, a significant improvement in performance is observed.
The point-wise relative error decreases from around $3\%$ in $S_4$ to around $0.1\%$ in $S_{32}$,
bringing it within the range of MC result uncertainties.

\begin{figure}[htbp]
\centering
\subfigure[$S_{4}$]
{\includegraphics[width=0.4\textwidth]{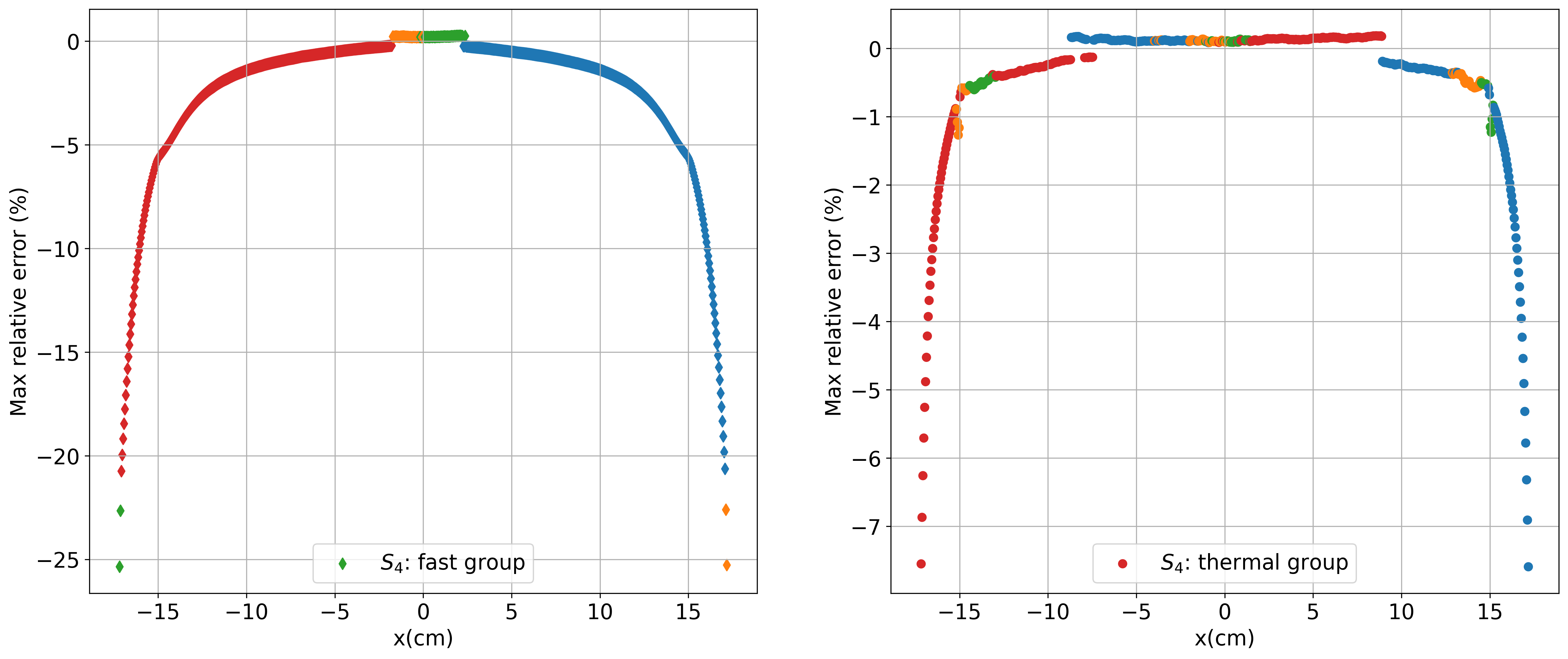}
  \label{fig::psi S4} 
} \hspace*{-0.9em}
\subfigure[$S_{8}$]
{\includegraphics[width=0.4\textwidth]{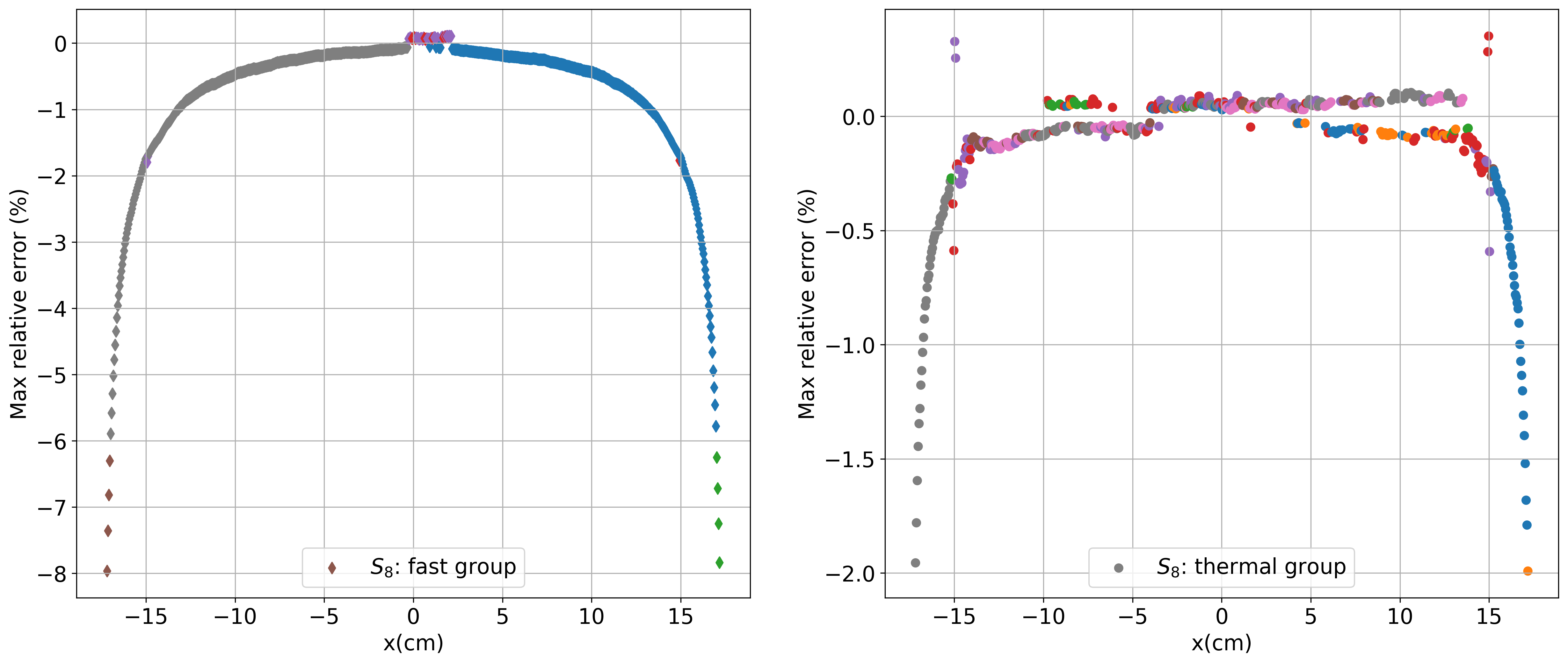}
  \label{fig::psi S8} 
} \hspace*{-0.9em}
\subfigure[$S_{16}$]
{\includegraphics[width=0.4\textwidth]{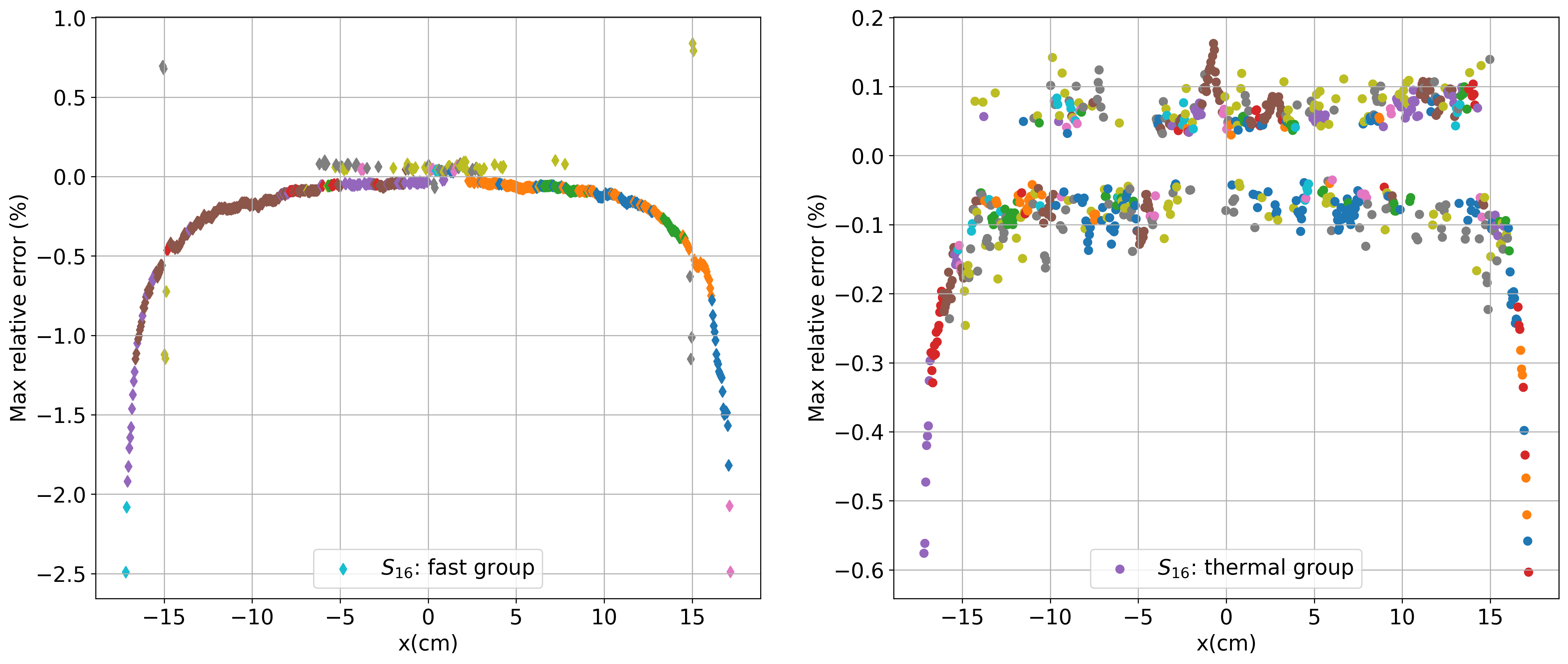}
  \label{fig::psi S16} 
} \hspace*{-0.9em}
\subfigure[$S_{32}$]
{\includegraphics[width=0.4\textwidth]{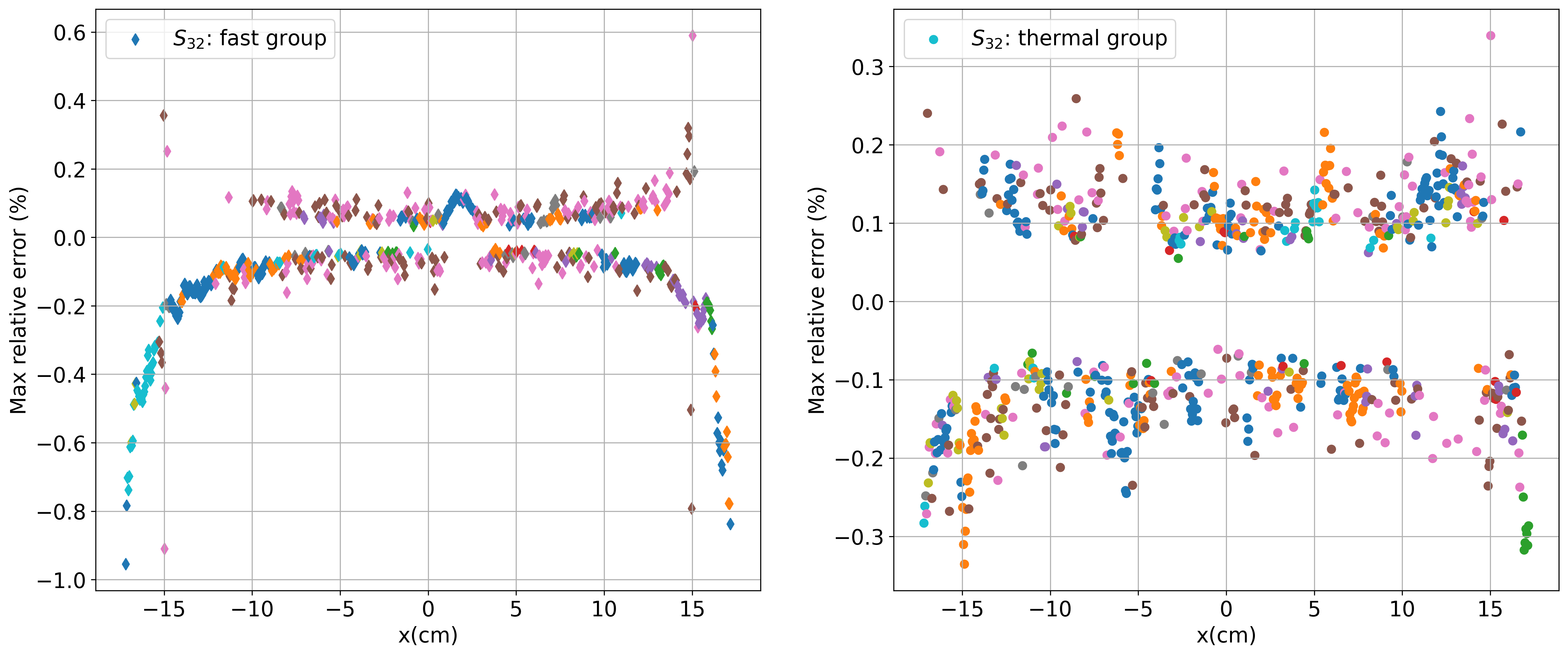}
  \label{fig::psi S32} 
} 

\caption{\scriptsize \SN angular fluxes compared with MC.
  The values plotted are the maximum relative error (\%) of angular flux for each group and each location.
The scatter colors are coded by the angles where the maximum error occurs.
}
\label{fig::psi comp}
\end{figure}

We observe similar patterns in the angular fluxes. In Fig.~\ref{fig::psi comp}, we present the maximum relative error among the discrete angles for each energy group and spatial position. Due to the discretization error with too few angles, we observe a maximum relative error of around $25\%$ for $S_4$ and $8\%$ for $S_8$, respectively. The maximum relative error decreases to around $1\%$ for $S_{32}$. Additionally, we note that for $S_4$ and $S_8$, where the $k_{eff}$ error is over $-9$pcm (see Table~\ref{tab::keff}), the maximum error is biased in the negative range. Conversely, when the $k_{eff}$ error decreases to below $1$pcm, the maximum error becomes symmetric about $0$. The errors are more prominent at the slab boundaries due to the reference values being close to 0, influenced by vacuum boundary conditions.

\section{Conclusions}
In this study, we showcased the treatment of scattering anisotropy using the analytical methods developed in our previous work for solving multigroup S$_N$ equations in slab geometry. For the slab problem derived from a typical pincell, we achieved -$47.9$pcm eigenvalue accuracy for the $S_4$ solution and less than $1$pcm eigenvalue accuracy in the $S_{32}$ solution. Notably, high accuracy was also observed in angular fluxes. As part of future work, we plan to extend the 1D solver to 3D neutron transport using schemes such as $2D-1D$ coupling and $3D$ nodal methods.


\section{Acknowledgments}
This work is supported by the Department of Nuclear Engineering, The Pennsylvania State University.


\scriptsize
\bibliographystyle{ans}
\bibliography{bibliography}
\end{document}